\documentclass[9pt, conference]{IEEEtran}
\IEEEoverridecommandlockouts
\usepackage{cite}
\usepackage{amsmath,amssymb,amsfonts}
\usepackage{algorithm}
\usepackage{algpseudocode}
\usepackage{multirow}
\usepackage{graphicx}
\usepackage{textcomp}
\usepackage{url}
\usepackage{scalerel}
\usepackage[dvipsnames]{xcolor}
\usepackage{fancyhdr}
\usepackage{booktabs}

\def\eqref#1{(\ref{#1})}
\def\BibTeX{{\rm B\kern-.05em{\sc i\kern-.025em b}\kern-.08em
    T\kern-.1667em\lower.7ex\hbox{E}\kern-.125emX}}

\usepackage{amsmath,amsfonts,bm}
\usepackage{mathrsfs}

\usepackage{amssymb,amsmath,amsthm}
\newtheorem{theorem}{Theorem}
\newtheorem{assumption}{Assumption}

\newtheorem{assumptionalt}{Assumption}[assumption]

\newenvironment{assumptionp}[1]{
  \renewcommand\theassumptionalt{#1}
  \assumptionalt
}{\endassumptionalt}

\newtheorem{innercustomthm}{Theorem}
\newenvironment{customthm}[1]
  {\renewcommand\theinnercustomthm{#1}\innercustomthm}
  {\endinnercustomthm}

\def\eqref#1{equation~\ref{#1}}

\def\1{\bm{1}}

\DeclareMathAlphabet{\mathsfit}{\encodingdefault}{\sfdefault}{m}{sl}
\SetMathAlphabet{\mathsfit}{bold}{\encodingdefault}{\sfdefault}{bx}{n}

\newcommand{\norm}[1]{\left\lVert#1\right\rVert}
\newcommand{\abs}[1]{\left|#1\right|}

\usepackage{tikz}

\usepackage{xcolor}

\definecolor{brightmaroon}{rgb}{0.76, 0.13, 0.28}
\definecolor{brown(web)}{rgb}{0.65, 0.16, 0.16}

\def\eqref#1{(\ref{#1})}

\def\eqref#1{(\ref{#1})}

\makeatletter 
\newcommand\footnoteref[1]{\protected@xdef\@thefnmark{\ref{#1}}\@footnotemark}
\makeatother

\begin{document}

\title{Latent Diffusion Bridges for Unsupervised Musical Audio Timbre Transfer  \\
}

\author{
\begin{tabular}{@{}c@{}}
Michele Mancusi$^{1,*}$\thanks{$*$ Equal contribution}
\qquad Yurii Halychanskyi$^{2, *, \dagger}$\thanks{$\dagger$ Work completed as a research intern at Sony}
\qquad Kin Wai Cheuk$^{3}$
\qquad Eloi Moliner$^{4}$\\
\qquad Chieh-Hsin Lai$^{3}$
\qquad Stefan Uhlich$^{1}$
\qquad Junghyun Koo$^{3}$
\qquad Marco A. Martínez-Ramírez$^{3}$\\
\qquad Wei-Hsiang Liao$^{3}$
\qquad Giorgio Fabbro$^{1}$
\qquad Yuki Mitsufuji$^{3}$
\end{tabular}
\IEEEauthorblockA{ \\
\\
$1$ Sony Europe B.V., Stuttgart, Germany \quad $2$ University of Illinois Urbana-Champaign, Urbana, IL, USA \\
$3$ Sony AI, Tokyo, Japan \quad $4$ Acoustics Lab, DICE, Aalto University, Espoo, Finland  \\
}
}



\maketitle

\begin{abstract}
Music timbre transfer is a challenging task that involves modifying the timbral characteristics of an audio signal while preserving its melodic structure. In this paper, we propose a novel method based on dual diffusion bridges, trained using the CocoChorales Dataset, which consists of unpaired monophonic single-instrument audio data. Each diffusion model is trained on a specific instrument with a Gaussian prior. During inference, a model is designated as the source model to map the input audio to its corresponding Gaussian prior, and another model is designated as the target model to reconstruct the target audio from this Gaussian prior, thereby facilitating timbre transfer. We compare our approach against existing unsupervised timbre transfer models such as VAEGAN and Gaussian Flow Bridges (GFB). Experimental results demonstrate that our method achieves both better Fréchet Audio Distance (FAD) and melody preservation, as reflected by lower pitch distances (DPD) compared to VAEGAN and GFB. Additionally, we discover that the noise level from the Gaussian prior, $\sigma$, can be adjusted to control the degree of melody preservation and amount of timbre transferred.
\end{abstract}


\section{Introduction}

Music timbre transfer refers to the process of altering the timbral characteristics of an audio signal, such as the type of musical instrument, while preserving other attributes like melody and rhythm. This technique has a wide range of applications, including audio editing, voice cloning~\cite{akuzawa2021conditional}. Various approaches have been developed for music audio timbre transfer. For example, methods based on variational autoencoders (VAEs)\cite{cifka2021self, bitton2020vector, bitton2018modulated, luo2022towards, ciranni2024cocola, luo2024dismix, luo2024unsupervised} learn latent representations for the audio. By modifying the timbre-related dimensions within this latent space and decoding the altered representation, the timbre of the original audio can be transformed. Generative Adversarial Networks (GANs)\cite{huang2018timbretron, kameoka2018stargan, 10096233, vaeGAN, kumar2019melgan} have also been widely explored for this purpose.
Recently, diffusion-based style transfer, popular in the image domain~\cite{su2022dual, zhou2023denoising, wang2023stylediffusion, he2024freestyle}, has been adapted to audio. However, text-to-audio style transfer methods often fail to retain musical content~\cite{copet2024simple, zhang2024instruct, zhang2024musicmagus, marianimulti, lin2024arrange, manor2024zero}. While optimal flow transport methods have shown better content preservation post-transfer, their application has been limited to tasks like declipping and dereverberation~\cite{moliner2023solving, GFB}.

In this paper, we explore the possibility of using optimal flow transport to transfer between two monophonic single instrument audio, for example, from violin to flute. Our method is based on dual diffusion bridges~\cite{su2022dual} which is a particular solution to the Schrödinger Bridge problem that aims to transport one data distribution into another~\cite{tang2024simplified,de2021diffusion,shi2024diffusion}. 
Our method offers several advantages over existing techniques: first, it does not require paired data,
allowing it to be trained on diverse datasets where only unpaired audio is available; second, each model is trained independently on a specific instrument. To add a new instrument, only one additional model needs to be trained, which can then be used in combination with existing models without retraining or finetuning. 
Popov et al. \cite{10094854} proposed a similar concept, achieving timbre transfer using a diffusion bridge in the mel-spectrogram domain. 
Their approach, however, requires gradient-guided inference with a set of control features to maintain content consistency.
 In contrast, our work is the first to perform timbre transfer using dual diffusion bridges without such requirements. 
Additionally, we also provide a theoretical explanation for the timbre transfer and cycle consistency, extending DDIB~\cite{su2022dual} to account for discretization and training errors.
The source code together with the supplementary material for timbre transfer using dual diffusion bridges is available online\footnote{\label{repo}\url{https://sony.github.io/diffusion-timbre-transfer/}}. 

\section{Methods}

\subsection{Inference and Training of the Diffusion Bridge}

%
Similar to Dual Diffusion Implicit Bridges (DDIB)~\cite{su2022dual}, our proposed method for music timbre transfer consists of a model trained on a source domain $D_\theta^{(s)}$, and a second model trained on a target domain $D_\theta^{(t)}$. A forward probability flow Ordinary Differential Equations (ODE)~\cite{song2020score} is used to convert data samples $x^{(s)}\in\mathbb{R}^d$ from the source domain distribution $p^{\text{src}}$ to the prior distribution $x^{(l)}\sim \mathcal{N}(\mathbf{0},\sigma_\text{max}^2 \mathbf{I})$, where $x^{(l)}\in\mathbb{R}^d$ has the same dimension as $x^{(s)}$, {and $\sigma_\text{max}$ is defined as the maximum noise level during the diffusion process}. 
Then, the reverse ODE is used to produce $x^{(t)}\in\mathbb{R}^d$ in the target domain distribution $p^{\text{tgt}}$ from $x^{(l)}$ via the same algorithm in reverse. The inference process is summarized as:  
\begin{equation}
\label{eq:ode}
\begin{split}
x^{(l)}&=\text{ODESolve}(x^{(s)};D_\theta^{(s)},\sigma_0,\sigma_{N-1}),  \\
x^{(t)}&=\text{ODESolve}(x^{(l)};D_\theta^{(t)},\sigma_{N-1},\sigma_0)
\end{split}
\end{equation}
{where $\sigma_{N-1}$ and $\sigma_0$ are the noise levels at step $N-1$ and step $0$}

In this work, we use the probablity flow ODE proposed by Karras et al.~\cite{karras2022elucidating}, and solve it using a 2nd order Heun numerical solver \cite{karras2022elucidating}, specified in Algorithm~\ref{algo:1}.
Conceptually, the algorithm increases deterministically the noise level of $x_0$ until it reaches the defined maximum $\sigma_\text{max}$. The reverse ODE can be obtained by reversing the order of $i$ in line 2.
In fact, any numerical method can be applied to solve Eq.~\eqref{eq:ode}, including the Runge–Kutta (RK) method~\cite{song2020score}, which generalizes Heun's as a special case.



Consider $x_i\in\mathbb{R}^d \sim \mathcal{N}(x_0, \sigma_i^2 \mathbf{I})$ an embedding corrupted with Gaussian noise with noise level $\sigma_i$ and dimension $d$.
 The models $D_\theta^{(s)}$ and $D_\theta^{(t)}$ are parameterized by deep neural networks and, given $x_i$ and the noise level $\sigma_i$ as input,
approximate a denoised estimate $\hat{x}_0$.
Formally, the denoiser models predicts the posterior expectation $D_\theta(\mathbf{x}_i, \sigma_i) =\hat{\mathbf{x}}_0 \approx \mathbb{E}[\mathbf{x}_0 | \mathbf{x}_i]$.

To train $D_\theta^{(s)}$, we minimize 
the $L_2$-norm 
between the original 
embedding $x_0$ and its denoised estimate $D_\theta^{(s)}(x_i;\sigma_i)$:
\begin{equation}
\mathbb{E}_{x_0 \sim p^\mathrm{src}, x_i \sim \mathcal{N}(x_0,\sigma_i^2\mathbf{I})} \left[\lambda(\sigma_i)|| x_{0}-D_\theta^{(s)}(x_{i};\sigma_{i}))||_{2}^2\right],
\label{eq:loss}
\end{equation}
where 
$\sigma_{ij}$ is the noise level for the noisy embedding $x_{ij}$ in a batch, and $\lambda(\sigma_i)$ is a weighting parameter depending on the noise level as specified in \cite{karras2022elucidating}. The same training objective applies for the target model $D_\theta^{(t)}$.


\begin{algorithm}
\caption{$\text{ODESolve}(x;D_\theta,\sigma_0,\sigma_{N-1})$}
\label{algo:1}
\begin{algorithmic}[1]
    \State $x_0=x$
    \For{$i \in \{0, \dots, N-1\}$}
        \State $d_i \gets (x_i - D_\theta({x}_i; \sigma_i))/{\sigma}_i$
        \State $x_{i+1} \gets {x}_i + (\sigma_{i+1} - \sigma_i)d_i$
        \If{$\sigma_{i+1} \neq 0$}
            \State $\hat{d}_i \gets (x_{i+1} - D_\theta(x_{i+1}; \sigma_{i+1}))/\sigma_{i+1}$
            \State $x_{i+1} \gets {x}_i + (\sigma_{i+1} - \sigma_i) \left(\frac{1}{2}d_i + \frac{1}{2}\hat{d}_i\right)$
        \EndIf
    \EndFor
    \State \textbf{return} $x_N$
\end{algorithmic}
\end{algorithm}

\subsection{Cycle Consistency}
DDIB shows that sequentially solving Probability Flow Ordinary Differential Equations (PF-ODEs) forms Schrödinger bridges and achieves \emph{cycle consistency}, where the process from source to latent to target and back to source recovers the original input. However, their analysis assumes perfect diffusion model training and no discretization errors in ODESolve, which is unrealistic. The argument of the following theorem extends DDIB's cycle consistency by accounting for ODESolve discretization errors using the RK method and training errors in diffusion models. It also provides a mathematical guarantee that our approach transfers the source distribution $p^{\text{src}}$ to the target distribution $p^{\text{tgt}}$ via Eq.~\eqref{eq:ode}.

\begin{theorem}[Distributional Cycle Consistency]\label{thm:main} Let $\hat{p}^{\text{tgt}}$ denote the density obtained by solving the ODEs in Eq.~\eqref{eq:ode} numerically via a $\kappa^{\text{th}}$-order RK method, starting from $x^{(s)} \sim p^{\text{src}}$. Let $h := \max_{i} \abs{\sigma_{i+1} - \sigma_{i}}$ be the discretization timestep.  Under certain assumptions, the total variation distance $\text{TV}$ between $\hat{p}^{\text{tgt}}$ and $p^{\text{tgt}}$ is bounded as: 
\begin{align*} 
\text{TV}\big(\hat{p}^{\text{tgt}}, p^{\text{tgt}}\big) \lesssim \mathcal{O}(\epsilon_{\text{DM}}) + \mathcal{O}(h^\kappa).
\end{align*}
Here, $\epsilon_{\text{DM}}$ represents the training error of diffusion models, and $\mathcal{O}(\cdot)$ and $\lesssim$ conceal a multiplication constant depending only on $p^{\text{src}}$ and $p^{\text{tgt}}$ and the numerical solver.
\end{theorem}
Let $\hat{p}^{\text{src}}$ denote the density obtained via a $\kappa^{\text{th}}$-order RK solver by solving the ``cycle manner'' ODEs as in DDIB's Proposition 3.1 (see Eqs.~\eqref{eq:cycle_ode_1} and \eqref{eq:cycle_ode_2} in the Appendix\footnoteref{repo}). A similar argument leads to the generalized cycle consistency property as in Theorem~\ref{thm:main}:
\begin{align*} 
\text{TV}\big(\hat{p}^{\text{src}}, p^{\text{src}}\big) \lesssim \mathcal{O}(\epsilon_{\text{DM}}) + \mathcal{O}(h^\kappa).
\end{align*}
We observe that a sample-wise bound can be derived by carefully analyzing the RK-solver but we do not pursue this complex study in this work.

\subsection{Model Architecture}

{
Following~\cite{karras2022elucidating}, we define our denoiser as:
\begin{equation}
D_\theta(x_i;\sigma_i)=c_\text{skip}(\sigma_i)x_i+c_\text{out}(\sigma_i)F_\theta(c_\text{in}(\sigma_i)x_i;c_\text{noise}(\sigma_i)),
\label{eq:denoiser}
\end{equation}
where the preconditioning parameters $c_\text{skip}(\sigma)$, $c_\text{out}(\sigma)$, $c_\text{in}(\sigma)$, and $c_\text{noise}(\sigma)$ are described in~\cite{karras2022elucidating}.
}
The model architecture $F_\theta$ is based on a one-dimensional U-Net architecture designed for processing latent representations derived from EnCodec~\cite{defossez2022highfi}, with an input channel size of 128 corresponding to the dimensionality of the EnCodec embeddings right after the encoder and before the quantization step. To ensure stable training the resulting EnCodec embeddings are normalized to have zero mean and unit variance for each channel using the precomputed statistics of the dataset. 
For more details, please refer to the source code\footnoteref{repo}.

\section{Experiments}\label{sec:exps}
\vspace{-1mm}
\subsection{Dataset}
We experiment with the CocoChorales Dataset~\cite{wu2022chamber},
which contains 
13 different solo instruments, from strings, woodwinds, and brass.
The data is sampled at 16 kHz . In this study, we train our models on violin (766.9 hours), flute (244.0), cello (466.2), and bassoon (230.3). For inference and evaluation, a subset of 1,000 audio samples per instrument is selected from the test set. Each audio file is either padded or truncated to a fixed duration of 17 seconds. We first resample the audio into 24 kHz. Subsequently, the audio data is transformed into a latent representation using EnCodec.

\subsection{Experimental Setup}

We trained separate models for each instrument.
The AdamW optimizer was used with a learning rate of $1.0 \times 10^{-4}$, 
$\beta_1 = 0.95$, $\beta_2 = 0.999$, and $\epsilon = 1.0 \times 10^{-6}$. We applied Exponential Moving Average (EMA) with $\beta = 0.995$ and a power factor of 0.7, and set weight decay to 0.001. Training was performed on a single Nvidia H100 GPU for 500 epochs with a batch size of 32.

Regarding inference, we experimented with different $\sigma_\text{max}$ and $\sigma_{N-1}$ parameters
, as detailed in Section~\ref{sec:results}.
We found that these parameters influence the audio quality and  melody preservation after timbre transfer. 
Additionally, we observed that the number of steps $N$ in line 2 of Algorithm~\ref{algo:1} is crucial for maintaining cycle consistency. For the main experiments, we used $N=$100 steps which is enough to maintain cycle consistency.

{
The scheduling method we used is based on \cite{karras2022elucidating}:
\begin{equation}
\sigma_{i} = \bigl(\sigma_{\text{min}}^{\frac{1}{\rho}} + \tfrac{i}{N-1} \bigl(\sigma_{\text{max}}^{\frac{1}{\rho}} - \sigma_{\text{min}}^{\frac{1}{\rho}}\bigr)\bigr)^\rho. 
\label{eq:sigma}
\end{equation}
During training, we sample $i$ uniformly from $i\sim \mathcal{U}(0,N-1)$ and compute the corresponding $\sigma_i$ using Eq.~(\ref{eq:sigma}). 
For inference, $i$ is discretized uniformly over the same range.
Unlike \cite{karras2022elucidating}, we apply the same schedule for both training and inference to simplify the hyperparameter search, as no significant performance difference was observed.
For all experiments, we set $\rho=9$, $\sigma_\text{min}=0.01$. To compute the preconditioning parameters in Eq.~(\ref{eq:denoiser}), we set $\sigma_\text{data}=1$. 
}\vspace{-1mm}
\subsection{Pitch-Shifting Augmentation}

To account for octave differences between $x^{(s)}$ and $x^{(t)}$ (e.g., flute to bassoon, cello to violin), we introduced pitch-shifting augmentation. Specifically, for the case where $x^{(s)}$ is flute, we randomly shifted the input pitch down by 1 to 25 semitones with a 35\% probability during training to match the pitch range of bassoon and cello. During inference, we experiment with shifting $x^{(s)}$ down -20 and -25 semitones to obtain $x^{(l)}$ using Eq.~(\ref{eq:ode}).

\subsection{Chunk-based Minibatch Optimal Transport Coupling}

We experiment with the technique proposed in the paper of Gaussian Flow Bridges (GFB)~\cite{GFB}, which improves content preservation and reduces gradient variance by minimizing trajectory curvature. While GFB only applied chunking to the time dimension of the waveforms, we experimented with chunking along the channel dimension on the EnCodec embeddings, using a time chunk size of 4 and a channel chunk size of 32.

\subsection{Evaluation}

To evaluate melody preservation, we use Basic Pitch~\cite{bittner2022lightweight} to transcribe the audio into an $(88, T)$ matrix representing the probability of a note being played, where $T$ is the time dimension, and 88 is the number of notes. We aggregate 88 notes into 12 pitch classes. Melody similarity between original and transferred versions is assessed using Jaccard distance~\cite{comanducci2024timbre} and Dynamic Pitch Distance (DPD), which incorporates Dynamic Time Warping (DTW) to account for temporal deviations.

To measure the quality of timbre transfer, we employed both objective and perceptual evaluation methods. We trained a simple neural network classifier on the EnCodec embeddings (averaged over the time dimension) and measured the accuracy. The classifier consists of two fully connected layers: the first layer maps the input dimension of 128 to 64 units, followed by a ReLU activation, and the second layer reduces the output to 5 units corresponding to the different timbre classes. Additionally, we utilized the Fréchet Audio Distance (FAD) computed on EnCodec embeddings, implemented in the Python library~\cite{fadtk}, to assess the perceptual quality of the generated audio. A listening test is also conducted where 20 participants are presented with 6 question to vote for the model that transfer to the target instrument better. 

\section{Results}\label{sec:results}

\subsection{Baseline Comparisons}
We compare our proposed method against an existing unsupervised timbre transfer model, VAEGAN~\cite{vaeGAN}. As shown in Table~\ref{tab:baseline}, VAEGAN achieves lower Jaccard distances, but it has a higher DPD than our method. Note that the Jaccard distance considers only the set of notes present in the audio while DPD consider also the temporal note alignment. Hence we conclude that our method preserves the musical structure of the melody better. We also observe that VAEGAN introduces significant artifacts in the generated audio, as evidenced by its higher FAD. These artifacts come from performing timbre transfer in the spectral domain, where a vocoder is required to convert spectrograms back into waveforms which introduces distortions. While using better vocoders could mitigate these artifacts, exploring such alternatives is beyond the scope of this paper. This highlights one of the advantages of our proposed model, which operates without the need for vocoders, thus avoiding these issues.

Next, we compare our method with a related approach, Gaussian Flow Bridges (GFB)~\cite{GFB}. GFB achieves a relatively low DPD, suggesting well preserved melodies. However, further analysis reveals that GFB fails to transfer $x^{(s)}$ to $x^{(t)}$ as reflected by the the near-zero timbre classification accuracy for both flute and violin. The FAD for flute-to-violin transfer is 60.8, close to the FAD between real flute and violin samples (59.8), suggesting minimal timbral change despite the transfer. This slight increase in FAD implies that GFB attempts to alter the audio but retains the original timbre. We attribute this limitation to GFB’s linear probability path, which seems to work only for tasks like declipping and dereverberation. In contrast, our method significantly outperforms GFB in timbre transfer while preserving the melodies. When the target instrument is in different octave range from the source (e.g., flute to bassoon), both VAEGAN and our method fail to preserve the melody. Techniques to address this will be discussed in the next section.


\begin{table}[]
\caption{Comparison with baseline models. Lower values indicate better performance for distance metrics (DPD, JD, and FAD).}
\centering
\label{tab:baseline}
\begin{tabular}{@{}cc|cc|cc@{}}
\toprule
\multirow{2}{*}{\textbf{Source - Target}}            & \multirow{2}{*}{\textbf{Model}} & \multicolumn{2}{c|}{\textbf{Pitch}} & \multicolumn{2}{c}{\textbf{Timbre}}                \\ 
                                  &                & \textbf{DPD}   $\downarrow$   &\textbf{JD}  $\downarrow$    & \textbf{FAD}  $\downarrow$   & \textbf{Acc.}  $\uparrow$                     \\ \midrule
\multirow{3}{*}{Violin - Flute}  & VAEGAN         & 0.32             &\textbf{0.05}   & 34.64          & 94.4\%                            \\
                                  & GFB            & \textbf{0.17}    & 0.09            & 65.61          & 0.0\%                             \\
                                  & proposed       & 0.24             & 0.08            & \textbf{24.83} & \textbf{97.6}\%  \\ \midrule
\multirow{3}{*}{Flute - Violin}  & VAEGAN         & 0.13             &\textbf{0.05}   & 60.77          & 85.1\%                            \\
                                  & GFB            & 0.11             & 0.11            & 75.30          & 0.1\%                             \\
                                  & proposed       & \textbf{0.08}    & 0.09            & \textbf{25.13} & \textbf{100.0}\% \\ \midrule
\multirow{3}{*}{Flute - Bassoon} & VAEGAN         & 1.92             & \textbf{0.12}            & 303.03         & 0.0\%                             \\
                                  & GFB            & 1.90             & 0.33            & 108.91         & 0.0\%                             \\
                                  & proposed       & \textbf{1.15}    & 0.15   & \textbf{33.01} & \textbf{100.0}\% \\ 
                                  \midrule
                                  
\end{tabular}
\vspace{-5mm}
\end{table}


Overall, our proposed method achieves a decent balance between timbre transfer quality and melody preservation compared to GFB and VAEGAN. This is supported by both pitch and timbre metrics. Additionally, the listening test results show that our model received 97 votes, while GFB and VAEGAN received 22 and 1 vote, respectively. Interestingly, in the few cases where listeners preferred GFB over our method, it was due to confusion between the timbres of the bassoon and cello. After excluding the two questions involving the bassoon and cello, the votes were 75 for our model, 4 for GFB, and 1 for VAEGAN.

The approach proposed by Popov et al. \cite{10094854} is not considered a baseline in our work due to the unavailability of the source code and insufficient details in the reported specifications. Moreover, their use of gradient-guided inference makes direct comparison with our method challenging.

\subsection{Effectiveness of Pitch-Shifting Augmentation}

Table~\ref{tab:pitch_shift} shows the results of pitch-shifting the source data $x^{(s)}$ to match with the pitch ranges of the target data $x^{(t)}$, specifically from flute to bassoon and cello. The results indicate a tradeoff between timbre quality and melody preservation. While the lower FAD without pitch-shifting indicates better audio quality, the high DPD reflects poor melody preservation. Shifting the flute down by 20 semitones improves melody preservation (lower DPD) but degrades audio quality (higher FAD). A further shift of 25 semitones down significantly worsens audio quality, as shown by the increased FAD. While a larger shift benefits melody preservation when the target is cello, it does not do so for bassoon. We hypothesize that excessive downsampling distorts $x^{(s)}$, which disrupts the bridging between $F_\theta^{(s)}$ and $F_\theta^{(t)}$. From this experiment, we conclude that a 20-semitone shift strikes an optimal balance between preserving timbre quality and maintaining melody integrity. Based on this result, we use 20-semitone shift for the case of flute-to-bassoon transfer in the next section when exploring different values of $\sigma_\text{max}$. 


\begin{table}[ht]
\caption{Effect of pitch shifting. The source instrument (flute) is in a different octave range than the target instruments.}
\label{tab:pitch_shift}
\centering
\begin{tabular}{@{}lc|cc|cc@{}}
\toprule
\multirow{2}{*}{\textbf{Target}} & \multirow{2}{*}{\textbf{Shift}} & \multicolumn{2}{c|}{\textbf{Pitch}} & \multicolumn{2}{c}{\textbf{Timbre}} \\
                                 &                                 & \textbf{DPD} $\downarrow$     & \textbf{JD} $\downarrow$    & \textbf{FAD}  $\downarrow$   & \textbf{Acc.} $\uparrow$   \\ \midrule
\multirow{3}{*}{Bassoon}         & 0                               & 1.15              & 0.15            &\textbf{ 33.01}            & \textbf{100.00\%}         \\
                                 & -20                             & \textbf{0.30 }             & \textbf{0.13 }           & 65.00            & 99.50\%          \\
                                 & -25                             & 0.48              & 0.19            & 104.38           & 88.20\%          \\ \midrule
\multirow{3}{*}{Cello}           & 0                               & 1.52              & 0.16            & \textbf{20.37}            & 99.80\%          \\
                                 & -20                             & 0.25              & 0.09            & 35.00            & \textbf{100.00\%}         \\
                                 & -25                             &\textbf{ 0.16}              &\textbf{ 0.08}            & 60.01            & 99.90\%          \\ \bottomrule
\end{tabular}
\vspace{-2mm}
\end{table}

\subsection{Impact of $\sigma_{\text{max}}$ on Timbre Transfer}

Table \ref{tab:sigma} explores the effects of different $\sigma_\text{max}$ (for training) and $\sigma_{\scaleto{N-1}{3pt}}$ (for inference) values. Starting with the case where $\sigma_{\scaleto{N-1}{3pt}} = \sigma_\text{max}$, models trained with $\sigma_{\text{max}} = 100$ in general show a better timbre quality (lower FAD) compared to those trained with $\sigma_{\text{max}} = 5$. This come in a price:  worse melody preservation (higher pitch distance). For the case of flute-to-violin, however, the FAD worsens when $\sigma_{\text{max}}$ is increased from 5 to 100. This suggests that some data domains such as violin might be sensitive to the value of $\sigma_{\text{max}}$ and that an optimal $\sigma_{\text{max}}$ might be required for different instruments. Despite this, we will use a unified $\sigma_{\text{max}}$ across all instruments in this work. The results indicate that lower $\sigma_\text{max}$ values favor melody preservation, while higher values enhance audio quality. Thus, $\sigma_{\text{max}}$ can serve as a control for balancing between melody preservation and timbre quality.

We also explore the possibility of sampling $x^{(l)}$ and $x^{(t)}$ with $\sigma_{\scaleto{N-1}{3pt}}<\sigma_\text{max}$ using Algorithm~\ref{algo:1} and Equation~(\ref{eq:ode}). Models trained with $\sigma_{\text{max}}=100$ and sampled with $\sigma_{\scaleto{N-1}{3pt}}=5$ perform similarly to the models trained and sampled with $\sigma_{\scaleto{N-1}{3pt}}=\sigma_{\text{max}} = 5$. Table~\ref{tab:sigma} also confirms the trade-off between audio quality and melody preservation when reducing $\sigma_{\scaleto{N-1}{3pt}}$ from $100$ to $5$ while keeping $\sigma_\text{max}=100$. Smaller $\sigma_{\scaleto{N-1}{3pt}}$ generally results in a better melody preservation. While setting $\sigma_{\scaleto{N-1}{3pt}}<\sigma_{\text{max}}$ lacks theoretical support in existing literature~\cite{song2020score,ho2020denoising,zhou2023denoising}, it is intriguing that this approach yields similar results as $\sigma_{\scaleto{N-1}{3pt}}=\sigma_{\text{max}} = 5$. These findings suggest that training with a sufficiently large $\sigma_{\text{max}}$  and inferring with with $\sigma_{\scaleto{N-1}{3pt}}<\sigma_{\text{max}}$ is a viable strategy.


\begin{table}[ht]
\caption{Impact of Different $\sigma$ Settings During Training and Inference for our proposed method.}
\label{tab:sigma}
\centering
\begin{tabular}{@{}llcc|cc|cc@{}}
\toprule
\multirow{2}{*}{\textbf{Source}} & \multirow{2}{*}{\textbf{Target}} & \multirow{2}{*}{$\sigma_{\text{max}}$} & \multirow{2}{*}{$\sigma_{\scaleto{N-1}{3pt}}$} & \multicolumn{2}{c|}{\textbf{Pitch}} & \multicolumn{2}{c}{\textbf{Timbre}} \\
                                 &                                  &                                        &                                                & \textbf{DPD} $\downarrow$      & \textbf{JD} $\downarrow$    & \textbf{FAD}  $\downarrow$   & \textbf{Acc.} $\uparrow$    \\ \midrule
\multirow{5}{*}{Violin}          & \multirow{5}{*}{Flute}           & 5                                      & 5                                              & \textbf{0.24}              & \textbf{0.08 }           & 24.83            & 97.6\%           \\
                                 &                                  & 100                                    & 100                                            & 1.13              & 0.29            & 19.39            & 98.5\%           \\
                                 &                                  & 100                                    & 50                                             & 1.04              & 0.27            & 16.77            & 98.7\%           \\
                                 &                                  & 100                                    & 20                                             & 0.69              & 0.19            & \textbf{13.47}            & \textbf{98.8\%}           \\
                                 &                                  & 100                                    & 5                                              & 0.30              & 0.10            & 20.26            & 97.8\%           \\ \midrule
\multirow{5}{*}{Flute}           & \multirow{5}{*}{Violin}          & 5                                      & 5                                              & 0.08              & 0.09            & \textbf{25.13 }           & \textbf{100.0}\%          \\
                                 &                                  & 100                                    & 100                                            & 0.97              & 0.27            & 37.79            & 99.7\%           \\
                                 &                                  & 100                                    & 50                                             & 0.89              & 0.25            & 37.42            & 99.9\%           \\
                                 &                                  & 100                                    & 20                                             & 0.47              & 0.16            & 27.62            & 100.0\%          \\
                                 &                                  & 100                                    & 5                                              & \textbf{0.07}              & \textbf{0.07}            & 25.41            & 100.0\%          \\ \midrule
\multirow{5}{*}{Flute}           & \multirow{5}{*}{Bassoon}             & 5                                      & 5                                              & 0.48              & 0.19            & 104.38           & 88.2\%           \\
                                 &                                  & 100                                    & 100                                            & 1.02              & 0.25            & 10.26            & 100.0\%          \\
                                 &                                  & 100                                    & 50                                             & 1.60              & 0.23            & \textbf{4.83 }            & \textbf{100.0}\%          \\
                                 &                                  & 100                                    & 20                                             & 1.51              & 0.23            & 7.71             & 100.0\%          \\
                                 &                                  & 100                                    & 5                                              & \textbf{0.40 }             &\textbf{ 0.14  }          & 65.45            & 78.5\%           \\ \bottomrule
\end{tabular}
\end{table}

\subsection{Results of Chunk-based Coupling Strategy} 

Since the previous results show that $\sigma_{\scaleto{N-1}{3pt}} = 5$ achieves a good balance in timbre transfer and melody preservation, only the case for $\sigma_{\scaleto{N-1}{3pt}} = 5$ and $100$ will be discussed. Table~\ref{tab:chunk} shows the effects of different chunk-based strategies. When $\sigma_{\scaleto{N-1}{3pt}} = 5$, chunking offers minimal benefits. In the flute-to-violin case, applying chunking to the time dimension fails to improve both pitch and timbre metrics. Even when chunking is applied to both time and channel dimensions, all metrics still perform worse than the models trained without chunking. Interestingly, timbre classification accuracy remains at 100\%. We attribute this to the complexity of the violin timbre, which is challenging to generate due to its distinct bow articulation, yet making it easier to classify. In the violin-to-flute transfer case, there is a slight improvement in melody preservation when chunking is used. In contrast, with $\sigma_{\scaleto{N-1}{3pt}}=100$, chunking leads to noticeable improvements, particularly in flute-to-violin transfers, where both FAD and melody preservation are significantly enhanced when using a time chunk size of 4 and a channel chunk size of 32. Violin-to-flute transfers also showed slight improvements. Despite these gains, the improvements were not sufficient to fully preserve the melody after transfer. Moreover, there is a tradeoff: while smaller chunk sizes tend to yield better results, increasing the number of chunks can lead to higher GPU memory usage.


\begin{table}[ht]
\caption{Effect of chunk-based coupling. (0,0): no chunking; (4,0): time chunking; (4,32) time and channel chunking.}
\label{tab:chunk}
\centering
\begin{tabular}{@{}lc|cc|cc@{}}
\toprule
\multirow{2}{*}{\textbf{Setting}}                                                                                   & \multirow{2}{*}{\textbf{Chunk}} & \multicolumn{2}{c|}{\textbf{Pitch}} & \multicolumn{2}{c}{\textbf{Timbre}} \\
                                                                                                           &                                 & \textbf{DPD} $\downarrow$      & \textbf{JD} $\downarrow$     & \textbf{FAD} $\downarrow$     & \textbf{Acc.} $\uparrow$    \\ \midrule
\multirow{3}{*}{\shortstack[l]{Flute to violin\\$\sigma_\text{max}=100$, $\sigma_{\scaleto{N-1}{3pt}}=5$}} & (0,0)                           & \textbf{0.07 }             & \textbf{0.07}            & \textbf{25.41}            & \textbf{100}\%            \\
                                                                                                           & (4,0)                           & 0.31              & 0.1             & 63.96            & 100\%            \\
                                                                                                           & (4,32)                          & 0.13              & 0.08            & 25.76            & 100\%            \\ \midrule
\multirow{3}{*}{\shortstack[l]{Violin to flute\\$\sigma_\text{max}=100$, $\sigma_{\scaleto{N-1}{3pt}}=5$}} & (0,0)                           & 0.3               & 0.1             & \textbf{20.26  }          & 97.8\%           \\
                                                                                                           & (4,0)                           & \textbf{0.2}               & \textbf{0.07}            & 23.19            & \textbf{99.2}\%           \\
                                                                                                           & (4,32)                          & 0.27              & 0.08            & 22.07            & 98.90\%          \\ \midrule
\multirow{3}{*}{\shortstack[l]{Flute to violin\\$\sigma_\text{max}=\sigma_{\scaleto{N-1}{3pt}}=100$}}      & (0,0)                           & 0.97              & 0.27            & 37.79            & 99.70\%          \\
                                                                                                           & (4,0)                           & 0.89              & 0.29            & 163.47           & 100\%            \\
                                                                                                           & (4,32)                          & \textbf{0.62 }             & \textbf{0.22}            & \textbf{30.25}            & \textbf{100}\%            \\ \midrule
\multirow{3}{*}{\shortstack[l]{Violin to flute\\$\sigma_\text{max}=\sigma_{\scaleto{N-1}{3pt}}=100$}}      & (0,0)                           & 1.13              & 0.29            & 19.39            & 98.5\%           \\
                                                                                                           & (4,0)                           & \textbf{0.93}              & \textbf{0.25 }           &\textbf{ 17.71  }          &\textbf{ 99.6\%}           \\
                                                                                                           & (4,32)                          & 0.96              & 0.26            & 23.08            & 97.60\%          \\ \bottomrule
\end{tabular}
\end{table}

\vspace{-2mm}
\subsection{Shared Latent Space and Cycle Consistency}

Equation~(\ref{eq:ode}) implies that all models $F_\theta^\diamond$ should point to the same prior distribution $\mathcal{N}(0,\sigma_\text{max})$. In this section, we explore the existence of shared latent space by directly sampling from the Gaussian $x^{(l)}\sim N(0,\sigma_{\scaleto{N-1}{3pt}})$ and use it to obtain both $x^{(s)}$ and $x^{(t)}$ via: 
\begin{equation}
\label{eq:shared_space}
x^{\diamond}=\text{ODESolve}(x^{(l)};D_\theta^{\diamond}, \sigma_{\scaleto{N-1}{3pt}},\sigma_0), \quad \diamond\in\{s,t\}\\
\end{equation}

If all models $F_\theta^{\diamond}$ share the same prior distribution, we expect $x^\diamond$ would share the same melody and change only in timbre depending on which model it is sampled from.  We study the case for violin and flute with $\sigma_{\text{max}}=100$. Out of 100 trials, we found that 49 of the $x^{(i)}$ pairs have similar melodic structure ($\text{DPD}<0.7$).
When using different $x^{(l)}$ for different $F_\theta^\diamond$, only 6 out of 100 trials produce similar melodic structures. These results indicate that $F_\theta^\diamond$ share the same prior distribution to some extent.


Cycle consistency is a critical aspect of audio-to-audio algorithms like timbre transfer. It is found to be sensitive to the number of sampling steps. As shown in the supplmentary material\footnoteref{repo}, the normalized L2 norm between embeddings decreased as the number of steps increased, improving reconstruction. However, this improvement comes at the cost of increased computational time, as more steps in the sampler require more processing time. This tradeoff highlights the need to balance reconstruction quality with efficiency in practice.

\section{Conclusion}
In this paper, we proposed using dual diffusion bridges for unsupervised musical timbre transfer. Through extensive experimentation, we demonstrated that our approach outperforms existing methods like VAEGAN and GFB in terms of timbre quality and melody preservation. We also provided a theoretical explanation for the timbre transfer and cycle consistency of our proposed method. The source code and the supplementary materials are available online\footnoteref{repo}.


\bibliographystyle{IEEEtran}
\bibliography{refs}

\clearpage
\newpage
\onecolumn
\section*{Appendix A. Theoretical Result and Its Proof}
\subsection*{A-1. Preliminaries }
 Let $p^{\text{src}}$ and $p^{\text{tgt}}$ be the source distribution and target distribution, respectively on $\mathbb{R}^d$ of dimension $d$. We consider the Ornstein-Uhlenbeck (OU) process\footnote{The statement and argument may be extended to a more general diffusion process. However, we leave it as a future work.} on an interval $[0,T]$ 
\begin{align}\label{eq:ou}
    dx_t = -x_t dt +\sqrt{2} dw_t, \quad x_0\sim p^{\diamond},
\end{align}
where $\diamond\in\{\text{src}, \text{tgt}\}$ and $\{w_t\}_{t\in[0,T]}$ is a Wiener process. We denote $p_t$ as the marginal density of the stochastic process $\{x_t\}_{t\in[0,T]}$ given by Eq.~\eqref{eq:ou}. It associates with the following PF-ODE~\cite{song2020score}
\begin{align}\label{eq:oracle_pf_ode}
    \frac{d}{dt}x_t = x_t + \nabla \log p_t^{\diamond}(x_t), \quad \text{where } \diamond\in\{\text{src}, \text{tgt}\}.
\end{align}
When $t=0$, it represents the clean data space (where $p^{\text{src}}$ and $p^{\text{tgt}}$ are supported on), and when $t=T$, it represents the latent noisy space.

For $\diamond\in\{\text{src}, \text{tgt}\}$, diffusion model $D_{\theta}^{\diamond}(x, t)$ is trained to approximate $\nabla \log p_t^{\diamond}(x)$ and leads to the following \emph{empirical PF-ODE}
\begin{align}\label{eq:emp_pf_ode}
    \frac{d}{dt}\hat x_t = \hat x_t + D_{\theta}^{\diamond}(\hat x_t, t).
\end{align}

Let $p$ and $q$ be two densities defined on $\mathbb{R}^d$. We define the \emph{total variation distance} between $p$ and $q$ as 
\begin{align*}
    \text{TV}(p, q):=\frac{1}{2}\int\abs{p(x)-q(x)}dx.
\end{align*}

Starting from $x^{\text{src}}\sim p^{\text{src}}$, the following ODEs solving  defines a \emph{cycle manner procedure}

\begin{equation}\label{eq:cycle_ode_1}
\begin{split}
\hat x^{\text{latent}}&=\text{ODESolve}(x^{\text{src}};D_\theta^{\text{src}},0,T),  \\
\hat x^{\text{tgt}}&=\text{ODESolve}(\hat x^{\text{latent}};D_\theta^{\text{tgt}},T,0),
\end{split}
\end{equation}
and then
\begin{equation}\label{eq:cycle_ode_2}
\begin{split}
\hat{\hat{x}}^{\text{latent}}&=\text{ODESolve}(\hat x^{\text{tgt}};D_\theta^{\text{tgt}},0,T),  \\
\hat{\hat{x}}^{\text{src}}&=\text{ODESolve}(\hat{\hat{x}}^{\text{latent}};D_\theta^{\text{src}},T,0),
\end{split}
\end{equation}

DDIB proves the \emph{cycle consistency property} that $\hat{\hat{x}}^{\text{src}} = x^{\text{src}}$, but assumes perfect diffusion model training and no ODE discretization errors, which are unrealistic. In Theorem~\ref{thm:main_rigorous}, we establish distributional cycle consistency by accounting for diffusion model training errors and ODESolve discretization errors.

\subsection*{A-2. Assumptions }
 We list up the assumptions which are mostly similar to those in \cite{huang2024convergence}. 

\begin{assumptionp}{A}[Compactly supported densities]\label{assum:compact_spt}
    Both $p^{\text{src}}$ and $p^{\text{tgt}}$ are compactly supported on a compact set in $\mathbb{R}^d$.
\end{assumptionp}

\begin{assumptionp}{B}[Training accuracy of diffusion model]\label{assum:dsm_error} Let $\epsilon_{\text{DM}}>0$. For $\diamond\in\{\text{src}, \text{tgt}\}$, 
\begin{align*}
    \int_{0}^{T} \mathbb{E}_{x_t\sim p_t(x)}\Big[\norm{D_{\theta}^{\diamond}(x_t, t)-\nabla \log p_t^{\diamond}(x_t)}_2^2\Big]dt \leq \epsilon_{\text{DM}}^2
\end{align*}
\end{assumptionp}

\begin{assumptionp}{C}[Smoothness of diffusion model]\label{assum:smooth} For $\diamond\in\{\text{src}, \text{tgt}\}$, assume that $D_{\theta}^\diamond(\cdot, t)$ is $\mathscr{C}^2(\mathbb{R}^d)$ for all $t\in[0,T]$. That is, it is twice continuously differentiable. Additionally, we assume that there is a constant $L_t>0$ so that 
\begin{align*}
    \norm{D_{\theta}^\diamond(\cdot, t)}_{\mathscr{C}^2(\mathbb{R}^d)}\leq L_t.
\end{align*}
We denote $L:= \int_{0}^{T} L_t dt$ and assume that $L<\infty$.
\end{assumptionp}

\subsection*{A-3. Full Statement of Theorem~\ref{thm:main} and Its Proof }
 To ensure precision, we slightly modify the notations used in the main manuscript. We present the theorem with time discretization, corresponding one-to-one with variance discretization~\cite{karras2022elucidating}. Let $t_{N-1}=T>\cdots>t_{i+1}>t_{i}>\cdots>t_0=0$ be the discretization timestep on $[0,T]$, and define $h := \max_{i\in\{0,\cdots,N-1\}} \abs{t_{i+1} - t_{i}}$. 

Starting from $x^{(s)} \sim p^{\text{src}}$, let $p^{\text{latent}}$ be the oracle density obtained by the forward-in-time PF-ODE (Eq.~\eqref{eq:oracle_pf_ode} with $\diamond=\text{src}$), and $\hat{p}^{\text{latent}}$ be the pushforward density obtained by solving the ODE (Eq.~\eqref{eq:emp_pf_ode} with $\diamond=\text{src}$) numerically:
\begin{equation*}
\begin{split}
\hat x^{(l)}&=\text{ODESolve}(x^{(s)};D_\theta^{\text{src}},0,T),  \quad x^{(s)} \sim p^{\text{src}}.
\end{split}
\end{equation*}
Now starting from the noisy latent space, let $\hat{p}^{\text{tgt}}$ be the density obtained by solving the ODE (Eq.~\eqref{eq:emp_pf_ode} with $\diamond=\text{tgt}$), starting from $\hat x^{(l)} \sim \hat{p}^{\text{latent}}$: 
\begin{equation*}
\label{eq:ode_time_forward}
\begin{split}
\hat x^{(t)}&=\text{ODESolve}(\hat x^{(l)};D_\theta^{\text{tgt}},T,0),  \quad \hat x^{(l)} \sim \hat{p}^{\text{latent}}.
\end{split}
\end{equation*}

We now present the full statement of Theorem~\ref{thm:main} along with its proof.
\begin{customthm}{1'}[Distributional Cycle Consistency]\label{thm:main_rigorous} Consider the ODE solvers are $\kappa^{\text{th}}$-order RK method. Under Assumptions~\ref{assum:compact_spt}, \ref{assum:dsm_error}, and \ref{assum:smooth}, the total variation distance $\text{TV}$ between $\hat{p}^{\text{tgt}}$ and $p^{\text{tgt}}$ is bounded by: 
\begin{align*} \text{TV}\big(\hat{p}^{\text{tgt}}, p^{\text{tgt}}\big) \lesssim \mathcal{O}(\epsilon_{\text{DM}}) + \mathcal{O}(h^\kappa). 
\end{align*}
Here, $\lesssim$ and $\mathcal{O}(\cdot)$ conceals a multiplication constant depending only on dimensionality $d$, $p^{\diamond}$ with $\diamond\in\{\text{src}, \text{tgt}\}$, and the pre-defined Runge–Kutta matrix~\cite{ixaru2004runge}.
\end{customthm}
\begin{proof}
    Applying \cite{huang2024convergence}'s Theorem~3.10 and its Remark~C.2 backward in time (from $T$ to $0$) to Eqs.~\eqref{eq:emp_pf_ode} and \eqref{eq:oracle_pf_ode} with $\diamond=\text{tgt}$, we obtain
\begin{align*} \text{TV}\big(\hat{p}^{\text{tgt}}, p^{\text{tgt}}\big) \lesssim \text{TV}\big(\hat{p}^{\text{latent}}, p^{\text{latent}}\big) + \mathcal{O}(\epsilon_{\text{DM}}) + \mathcal{O}(h^\kappa). 
\end{align*}
    Now applying the same theorem but forward in time (from $0$ to $T$) to Eqs.~\eqref{eq:emp_pf_ode} and \eqref{eq:oracle_pf_ode} with $\diamond=\text{src}$, we obtain
\begin{align*}
    \text{TV}\big(\hat{p}^{\text{latent}}, p^{\text{latent}}\big) \lesssim  \mathcal{O}(\epsilon_{\text{DM}}) + \mathcal{O}(h^\kappa),
\end{align*}
    as we start from the same initial distribution $p^{\text{src}}$. Combining these two inequalities, we derive the desired bound:
\begin{align*} \text{TV}\big(\hat{p}^{\text{tgt}}, p^{\text{tgt}}\big) \lesssim \mathcal{O}(\epsilon_{\text{DM}}) + \mathcal{O}(h^\kappa). 
\end{align*}
\end{proof}

We note that a sample-wise bound (instead of a distributional bound) can also be derived by analyzing the RK-solver in detail. Additionally, the bounds in Theorem~\ref{thm:main_rigorous} can be further refined using advanced techniques, but we do not pursue this overly complex mathematical analysis in this work.

\end{document}